\documentclass[twocolumn,noshowpacs,preprintnumbers,amsmath,amssymb,showkeys,nofootinbib]{revtex4}

\usepackage{graphicx}	
\usepackage{dcolumn}	
\usepackage{bm}		

\begin{document}

\preprint{Draft}
\title{Temperature Dependence of Light Absorption by Water}
\author{J.B.Cumming}
\email{cumming@bnl.gov}
\affiliation{Chemistry Department, Brookhaven National Laboratory,
Upton, New York 11973}

\date{\today}

\begin{abstract}
A model is described that relates the temperature coefficient
of the optical absorption spectrum of pure water to the
frequency derivative of that spectrum and two parameters that
quantify the dependence of a peak's amplitude and its position on temperature.
When applied to experimental temperature coefficients, it provides
a better understanding of the process than the analysis currently in use.
\end{abstract}
\keywords{water, optical absorption,
temperature coefficient}
\maketitle
\section{Introduction}
Understanding the temperature dependence of optical absorption takes on increasing importance
as more accurate spectra become available. A high quality absorption spectrum
from the infrared to the blue for water at $22^\circ$C
can be obtained by combining the data of
Kou, Labrie, and Chylek\cite{Kou93} with that of Pope and Fry \cite{Pop97}.
Kou \textit{et al.} studied the range
from 666 to 2400~nm with a Fourier-transform infrared spectrometer.
They derived absorption coefficients from
the ratio of transmittances obtained with different path lengths.
This exactly cancels reflection losses at the air-window and window-water
interfaces and absorption by the cell windows. Only transmittance values
between $20\%$ and $60\%$ were used to avoid saturation and base line
uncertainties. Absorption coefficients from 380 to 727.5 nm were obtained by
Pope and Fry using a novel integrating cavity absorption meter(ICAM). In the
ICAM, optical power loss from a water-filled highly-reflective cavity is 
measured as a function of wavelength. Scattering, which is important at the lower
end of this wavelength range, does not cause a loss of power, hence the ICAM
measures absorption.

The absorption coefficient of water decreases by some six decades from the
infrared through the blue. The infrared spectra in Fig.\,1 from Kou \textit{et al.}\cite{Kou93}
show a general exponential decrease vs frequency that is modulated by
various peaks, valleys, and shoulders indicating absorption by discrete states or
groups of states. The amplitude of these features decreases relative
to the background continuum with increasing frequency.
The use of frequency (in wavenumbers) emphasizes the approximately
uniform spacing of these features.
%
\begin{figure}
\includegraphics[width=3.5in]{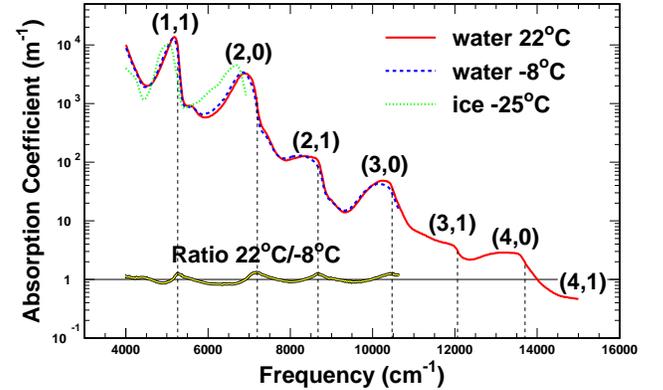}
\caption{\label{fig:fig1} Absorption spectra of water,
supercooled water, and ice from Kou, Labrie, and Chylek\cite{Kou93}.
Oscillator assignments are shown in parentheses above the peaks, see text. 
The ratio of the 22$^\circ$ spectrum to the -8$^\circ$ spectrum
is shown in the lower part of the figure. Vertical dashed lines mark
inflection points on the high frequency side of each 22$^\circ$ peak.} 
\end{figure}
%

The vibrational level spectrum of an isolated water molecule can be described in terms of
harmonics or combinations of three normal modes: a symmetric and an asymmetric stretch,
and a bending or scissors mode. Single levels are broadened by rotational structure.
In condensed phases, formation and breaking of hydrogen bonds further broadens levels
and may introduce additional modes. The result is not a continuum for
low excitations: Spectra in Fig.\,1 retain various discrete features.
Patel and Tam \cite{Pat79} have identified absorption peaks in the liquid with the
band structure of two anharmonic oscillators. The first was based on a single
(degenerate) stretching mode.
The second, a parallel series of states, was upshifted by a bending excitation.
We will refer to these as the stretching band and the combination band, respectively.
The frequency in wavenumbers of levels in terms of stretching and bending numbers,
($n_s,n_b$), is given by:
\begin{eqnarray}	
 {\nu (n_s,0)} & = & {n_s(3620 - 63n_s)}  \\
 {\nu (n_s,1)} & = & {\nu (n_s,0) + 1645}. 
 \nonumber
\end{eqnarray} 
The peaks and shoulders in Fig.\,1 are labeled with their $(n_s,n_b)$ assignments.

The $22 ^\circ$C absorption spectrum of Pope and Fry \cite{Pop97} at has a
minimum in the vicinity of 420 nm.
Recent data from Abe and the Super-Kamiokande Collaboration\cite{Abe11} at $13 ^\circ$C, 
and from Griskevich, Renshaw, and Smy\cite{Gri12} at $15 ^\circ$C, suggest that this
minimum is appreciably deeper and shifted to a higher frequency.
The present paper is an outgrowth of a study of scattering and
absorption by water in the vicinity of the minimum. When comparing
various measurements, it was desirable to separate the influence of
temperature from those due to other systematic effects. The classical treatment
of temperature dependence is to fit the absorption spectrum to an appropriate
set of Gaussians and then to determine a unique temperature coefficient for each peak
from experimental data. An example of this sort of analysis due to
Jonasz and Fournier \cite{Jon07} is described in Sec.\,3.
Measurements of Buiteveld, Hakvoort, and Donze \cite{Bui94},
of Pegau, Gray, and Zaneveld \cite{Peg97}, and of
Langford, McKinley, and Quickenden \cite{Lan01} indicate that absorption peaks in the visible
not only shift upward, but also move to higher frequencies with increasing temperature.
The model developed below predicts that shifts of peak position may result in
apparent positive and negative temperature coefficients for a given peak. 
A variety of experimental observations are a natural consequence of this model.  
\section{The Model}
Ratios of absorption coefficients at $22^\circ$C to those at $-8^\circ$C are
plotted in the lower part of Fig.\,1. Although the individual absorption spectra
show considerable structure, the ratio takes a simple form; 
a series of cusps where the ratio is greater than one, separated by regions where
it drops below unity. Vertical dashed lines in Fig. 1 indicate that the cusps
correspond with inflection points on the high frequency sides of peaks
in the $22^\circ$C spectrum, not their centers.
This suggested a connection between temperature
dependence and the shape of the absorption spectrum.

Consider part of an absorption spectrum, $a(\nu,T)$, in the vicinity of some
spectral feature located at frequency $\nu_0$ where the absorption coefficient
is $a_0$. We assume that $a$ is separable:
\begin{equation}	
		a(\nu,T) = a_0(T) f(\nu -\nu_0(T)).
\end{equation}
Differentiating Eq. 2 with respect to $T$ and substituting $df/d\nu_0 = -df/d\nu$
leads to:  
\begin{equation} 
\frac{1}{a} \frac{da}{dT} = \frac{1}{a_0} \frac{da_0}{dT}
                          - \frac{1}{a}  \frac{da}{d\nu} \frac{d\nu_0}{dT}. 
\end{equation}
The temperature coefficient\footnote{We will use ``temperature coefficient''
for the fractional temperature
coefficient, reserving ``absolute temperature coefficient'' for $da/dT$.}  
is the sum of two terms. The first is the trivial ``classical'' interpretation:
If a temperature change of $+1^\circ$C increases the intensity of
an absorbing peak by $+1\%$, then
the temperature coefficient is $+1\%$ per $^\circ$C across the whole peak. 

The second term on the right side of Eq.\,3 has more interesting properties.
If the region of interest contains a peak, ${da}/{d\nu}$ is zero
at the peak position.
The temperature coefficient will change sign in scanning across that peak. 
That maximum temperature changes match the inflection points on the high frequency
side of the peaks in Fig.\,1, indicates that $d\nu_0/dT$ is positive in the infrared.
Experimental data presented in Sec.\,3 indicate this is true in the visible as well. 
As a consequence, temperature coefficients will be negative on the low
frequency side of a peak and positive on the high side.  

If the region of interest is a decreasing continuum, the derivative is
negative throughout, and temperature coefficients will be positive. Consider then the
progressive transition from a spectral region dominated by individual peaks
to one of a decreasing continuum. In the first region, temperature coefficients should
show a bipolar oscillation, with some upwards bias due to the underlying continuum and any contribution from the classical term of Eq.\,3. Moving up in frequency,
the relative contribution of peaks decreases and that of the continuum increases.
The amplitude of the oscillations will decrease, and the upward bias will increase.
At some point, the negative lobes will be balanced out by the bias, and an absorption
peak will appear to have only a positive temperature coefficient peak.
However, this will not be centered at the
peak position but on the inflection point on its high frequency side.
With further increases in frequency, the envelopes of the upper and lower coefficient       
oscillations will converge to the continuum value.

\section{Experimental Data}
The logarithmic frequency derivative, $(1/a)(da/d\nu)$, is compared with
the fractional temperature coefficient, $(1/a)(da/dT)$, in Fig.\,2.  
The derivative of the $22^\circ$C spectrum in Fig.\,1 is shown as a hollow curve in Fig.\,2(a).
The solid curve is from the absorption spectrum of Pope and Fry\cite{Pop97},
also at $22^\circ$C.
Peak positions of the stretching and combination band
absorptions are shown by $(n_s,n_b)$ labels.
Each of the
oscillator states is marked by a sharp negative derivative spike on its high frequency side.
Positive excursions are less pronounced due to, in part, the overall decreasing trend of the spectrum. 
%
\begin{figure}
\includegraphics[width=3.5in]{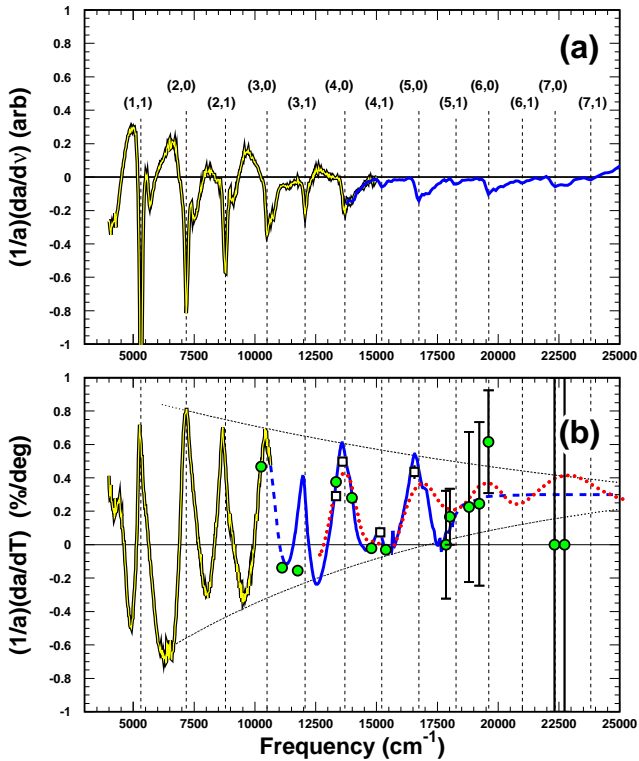}
\caption{\label{fig:fig2} Comparison of the frequency derivative
of the water absorption spectrum at $22^\circ$C with the temperature coefficient
of that spectrum. Hollow curves in (a) and (b) are from
Kou \textit{et al.}\cite{Kou93}. The solid curve in (a) is
from Pope and Fry\cite{Pop97}. The solid curve in (b) is based on the 
measurements of Langford \textit{et al.}\cite{Lan01}. A dashed curve connects it to
the hollow curve.
Circles show experimental values from Pegau \textit{et al.}\cite{Peg97},
squares those from Trabjerg and Hojerslev \cite{Tra96}.
The heavy dotted curve is from an analysis by Jonasz and Fournier\cite{Jon07}. 
Thin dotted curves show the approximate envelopes of the temperature
coefficient oscillations. Positions of frequency derivative minima
are shown by vertical thin dashed lines.} 
\end{figure}
%
These derivative oscillations damp down in moving into the visible.
Negative excursions corresponding to each state persist, but the
positive excursions transform into a small negative background
between the larger negative excursions. Pronounced peaks and valleys have given way to
an exponentially decreasing absorption spectrum
where specific absorbing states appear only
as shoulders. Note that the amplitude of negative derivative peaks associated
with $(n_s,1)$ transitions decreases more rapidly than those for $(n_s,0)$.
They are decidedly less pronounced in the visible.
This suggests that the dependence on frequency is different for
the two bands. Positions of the derivative minima
are projected on to Fig.\,2(b) by vertical dashed lines.  

The dependence of the temperature coefficient
on frequency is shown in Fig.\,2(b).
The hollow curve was obtained from the data of Kou \textit{et al.}\cite{Kou93}.
The solid curve is based on the extensive measurements of
Langford, McKinley, and Quickenden \cite{Lan01}. 
These authors studied the temperature dependence of
decadic absorptivity, $A$, over the wavelength
range from 550 to 900 nm and for temperatures between $15$ and $60^\circ$C.   
At any wavelength, $A$ was observed to vary linearly with $T$.
Their published decadic $dA/dT$ values were converted to $da/dT$, then to
$(1/a)(da/dT)$ using the absorption coefficients of
Kou \textit{et al.} and of Pope and Fry. A spline interpolation joining
the two measurements is shown as a heavy dashed line in Fig.\,2(b). 

Buiteveld \textit{et al.}\cite{Bui94} have
derived $da/dT$ values for wavelengths from 400 to 800 nm (12,500 to 25,000 cm$^{-1}$) 
from absorption spectra obtained between $2.5$ and $45^\circ$C.
They noted that $da/dT$ was nearly constant
at $0.0012$ m$^{-1}$ per $^\circ$C below 580 nm.
Pegau \textit{et al.} have concluded
there was a systematic bias of $0.0011$ in that value.
We subtracted that bias prior to converting the Buiteveld \textit{et al.}
$da/dT$ values to $(1/a)(da/dT)$. These were in agreement those of
Langford \textit{et al.} for wavelengths longer than
580 nm. They have not been included in Fig.\,2(b) because of
uncertainties due to the baseline correction,

Absolute temperature coefficients
from Pegau \textit{et al.}\cite{Peg97} were converted to the $(1/a)(da/dT)$ values which
are shown as circles in Fig.\,2(b). With the exception of the point at 850 nm (11,760 cm$^{-1}$),
there is good agreement with the curve based on Kou \textit{et al.} and Langford \textit{et al.}.   
Four points due to Trabjerg and Hojerslev \cite{Tra96},
squares in Fig.\,2(b), are also in agreement with that curve.
 
Jonasz and Fournier\cite{Jon07} have reported a detailed analysis of the absorption
spectrum of water and its temperature coefficient in the range from 380 to 800 nm
The absorption spectrum could
not be reproduced with a ten-member basis-set of Gaussians
corresponding to the stretching and combination band states $(4,0)$ to $(8,1)$.
Five additional Gaussians were needed to obtain
``impressive'' agreement with the experimental spectrum.
Jonasz and Fournier then used available data to derive fractional temperature
coefficients for each of the 15 Gaussians.
This classical approach is equivalent to setting
$d\nu_0/dT = 0$ in Eq.\,3. Their derived spectrum of temperature coefficients,
the dotted curve in Fig.\,2(b), is a reasonable fit to the experimental data
for frequencies below 15,500 cm$^{-1}$ where the extra Gaussians were added.
In particular, it reproduces the negative values to the left of the $(4,0)$ peak and the 
near zero values in the vicinity of the $(4,1)$ position.   
Agreement is poorer at higher frequencies. Oscillations of their curve in that region
are a consequence of their ascribing temperature coefficients
of $0.45\%$\,per\,$^\circ$C to the $(n_s,0)$ absorptions
and $0.20\%$\,per\,$^\circ$C to the $(n_s,1)$ absorptions in their original ten-member basis set.
The oscillations would have continued down
to 12,500 cm$^{-1}$ (800 nm) were it not for the five extra Gaussians.
One positioned near the $(4,0)$ peak had a larger positive coefficient. The
other four located between the $(4,0)$ and $(5,0)$ peaks had large negative values. The physical
significance of unusual properties of these phantom states is not clear.  

Several features of Fig.\,2 support the derivative model. The one-to-one correspondence
between frequency derivative minima and temperature coefficient maxima is striking
and implies that $d\nu_0/dT$ in Eq.\,3 is positive. The nearly symmetric, two-lobed
form of the temperature coefficient spectrum in the infrared provides strong evidence
that frequency shift is a major contributor to temperature dependence. The region
of near zero temperature coefficients at $\sim 15,000$ cm$^{-1}$ is interpreted as being due
to a balance between the negative lobe of the $(5,0)$ absorption and and the positive
contribution from the decreasing continuum. The $(4,1)$ absorption makes a small
contribution in this valley.   

Approximate upper and lower envelopes of the temperature coefficient oscillations
for the $(n_s,0)$ absorptions are shown as thin dashed curves in Fig.\,2(b). The damping of
oscillations with increasing frequency, and the 
the approach to a limit of $\sim +0.3\%$\,per\,$^\circ$C which is shown as a heavy dashed line,
are consistent with the change from discrete absorption peaks
in the infrared to a decreasing continuum in the visible. Note that the $0.3\%$\,per\,$^\circ$C 
applies to absorption only.
Because scattering has little if any temperature dependence,
the fractional temperature coefficient for attenuation is reduced by the
ratio of absorption to attenuation. For example, scattering and absorption coefficients
are equal at 425 nm \cite{Abe11} and the fractional temperature coefficient for attenuation
is then $0.15\%$\,per\,$^{\circ}$C.      
\section{Conclusions}
The temperature dependence of optical absorption by water has a complex dependence
on frequency related to properties of the absorbing states or continuum.
The present model, which connects the
temperature dependence to the frequency derivative of the
absorption spectrum, provides a framework for understanding this dependence. 
The composite curve in Fig.\,2(b) based on Kou \textit{et al.} and Langford \textit{et al.} 
with the dashed interpolation and extrapolation, provides practical temperature corrections
for absorption coefficients.
The model suggests that analyses of the type which fit an absorption
spectrum with Gaussians should use the derivative of the Gaussians to fit the temperature
coefficient spectrum. A continuum background must be included in  both.
\begin{acknowledgments}
The author wishes to thank M.\,Smy for communicating his unpublished results.
This work, conducted at Brookhaven National Laboratory, was supported by the 
U.S. Department of Energy's Offices of Nuclear Physics and High Energy Physics,
under Contract DE-AC02-98CH10886.
\end{acknowledgments}
%
%
%

%
\end{document}